# Adaptive Fuzzy Controller for Synchronous Generator

**Ioan Filip, Octavian Proştean, Cristian Vaşar, Iosif Szeidert**

Department of Automation and Applied Informatics, Faculty of Automation and Computer Sciences, "Politehnica" University from Timisoara, Av. V. Parvan, No.2, 300223, Timisoara, Romania, Phone: (0040) 256 403237, Fax: (0040) 256 403214, ifilip@aut.utt.ro

*Abstract: This paper describes a PI fuzzy adaptive control structure. Based on classical PI fuzzy control structure (with integration on controller output) the authors have developed, implemented and validated an on-line self-tuning mechanism of regulator's parameter. A study case for validation of proposed tuning mechanism is presented and analyzed with application to control of synchronous generator excitation system.*

## 1 Introduction

The complexity rising of the controlled process, doubled by the stringent necessity of assuring enhanced performance quality indicators for the controlled process, sustained by the great development of computer technology, imposes fuzzy control strategies as modern solutions in the domain of industrial control applications.

Lotfi A. Zadeh introduced the concept of fuzzy logic through a paper presented in 1965 at the Berkeley University from California. Later, in 1973, Zadeh elaborated a new paper in which appears the concept of "linguistic variables" associated to a fuzzy set. It was followed by other researches, first industrial application based on fuzzy logic being a cement furnace made in 1975 in Denmark. In Japan, Seiji Yasunobu and Soji Miyamoto from the Hitachi Company demonstrated by simulation (in 1985) the superiority of the controlled fuzzy systems in the field of railway transportation. Their idea was adopted and implemented in 1987 by the Sendai transportation firm. The fuzzy logic becomes a vogue in Japan and started the development of industrial applications. In the evolution of the fuzzy control systems there can be identified two stages: a stage of the developments preponderant theoretical (starting from the middle of 60's and which continues nowadays), respectively a stage of practical implementations (starting with the 80's) quickened by the explosive development of microelectronics and the computer science, and put it across by the apparition on the market of some commercial products offered by well-known firms. The technologic development had lead to a new revival of the scientific and applicative researches in this field.

The researches presented in this paper regarding a fuzzy control strategy dedicated to the control of the energetic systems, in which the existent process presents a particular complexity (including the case of the synchronous generator) are relative well presented in the specialty literature. The number of industrial implementations in the energetic field is relatively reduced. The classic solutions of control, already tested for years and with high degrees of trust, are yet prevailing in an overwhelming proportion. The control of synchronous generator excitation system is one of the most used techniques to assure power system stability and desired functioning for the entire power system.

The performed study cases with a PI or a PID fuzzy control structure have highlighted the necessity of on-line re-tuning of those controller's parameter in the case of processes with different dynamic behaviors related to the functioning regime. These kinds of processes are represented by power systems, exemplified by a synchronous generator. In figure 1 is shown the generalized structure of an adaptive fuzzy control system. [1][3]

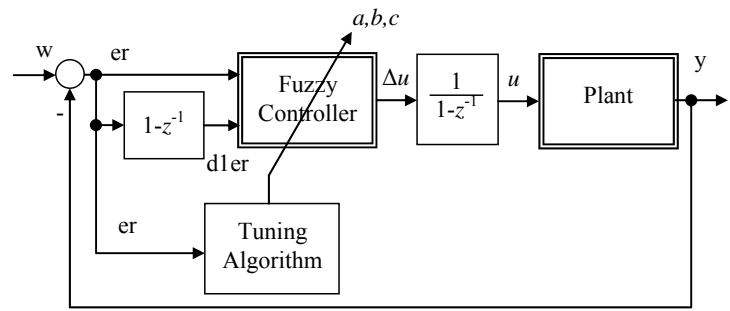

Figure 1

Adaptive PI fuzzy control structure

## 2   Model of the Process - Synchronous Generator

The synchronous machine (considered as a plant), mostly appearing as synchronous generator, represents the main installation in major electric power systems. Due to its active role within the system - being used to supply electric power and to modify the voltage and the circulation of active and reactive power – the synchronous generator shows a capital significance for designers and engineers involved in solving system problems.

Beginning from the Park's equations, the non-liner model of a synchronous generator connected to an infinite bus through a long transmission line (Figure 2) is described by the next ten equations depicted as follows [6]:

$$\frac{d}{dt}(\delta) = \omega_0 s \tag{1}$$

$$M\frac{d}{dt}(s) = -k_d s + T_m - T_e \tag{2}$$

$$T'_{d0}\frac{d}{dt}(e'_q) = v_f - (x_d - x'_d)i'_d - e'_q \tag{3}$$

$$T''_{d0}\frac{d}{dt}(e''_q) = e'_q - (x'_d - x''_d)i_d - e''_q \tag{4}$$

$$T''_{q0}\frac{d}{dt}(e''_d) = (x_q - x''_q)i_q - e''_d \tag{5}$$

$$e''_d = v_d + \gamma_a i_d - x''_q i_q \tag{6}$$

$$e''_q = v_q + \gamma_a i_q + x''_d i_d \tag{7}$$

$$T_e = e''_d i_d + e''_q i_q - \left(x''_d - x''_q\right)i_d i_q \tag{8}$$

$$v_t^2 = v_d^2 + v_q^2 \tag{9}$$

$$T_{ex}\frac{d}{dt}(v_f) = u - v_f \tag{10}$$

$$v_d = v_b \sin\delta + \gamma_e i_d - x_e i_q \tag{11}$$

$$v_q = v_b \cos\delta + \gamma_e i_q + x_e i_d \tag{12}$$

where: $T_{do}', T_{do}'', T_{qo}''$ - time constants; $T_{ex}$ - excitation time constant; $\delta$ - rotor angle; $\omega_0$ - synchronous speed; $Tm, Te$ - electric and mechanical torque; $i_d, i_q$ - direct and quadrature axis currents; $k_d$ - damping coefficient; $M$ – inertia; $e_q'$ - quadrature axis transient voltage; $e_d''$, $e_q''$ - direct and quadrature axis supratransient voltages; $v_d, v_q$ - direct and quadrature voltages; $v_f$ - field voltage; $v_t$ - terminal voltage (output); $v_b$ - bus voltage; $u$ - control input of excitation voltage; $x_d, x_q$ - direct and quadrature axis reactances; $x_d', x_q'$ - transient reactances; $x_d'', x_q''$ - subtransient reactances; $x_e$ - transmission system reactance; $\gamma_a$ - armature resistance; $\gamma_e$ - transmission system resistance; $x_e$ - transmission system reactance; $s$ – slip;

In this paper the considered synchronous generator is connected to a power system and has a local consumer at the generator terminals (Figure 2).

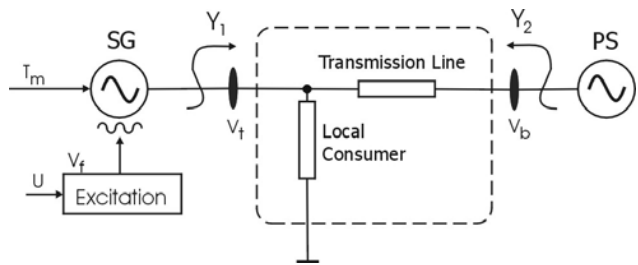

Figure 2

Synchronous generator connected to a power system

In the Figure 2 the following notations are used: SG – synchronous generator, PS- power system, $T_m$ – mechanical torque, $Y_1=G_1+jB_1$ – transmission network admittance to the SG terminals, $Y_2=G_2+jB_2$ – transmission network SG-PS ($G_1$, $G_2$ –conductatances, $B_1$, $B_2$ –susceptances).

The generalisation of the considered structure consists in introduction in the connection network, beside the transmission line, of a consumer connected at generator terminal [5][6]. The considered connection network can be treated as a quadripole (Figure 2).

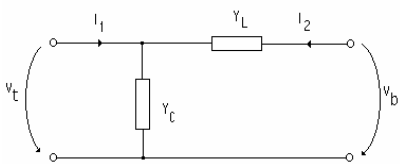

Figure 3

Equivalent connection network of synchronous generator to a power system

This quadripole (figure 3) is characterised by the following matrix equation:

$$\begin{bmatrix} I_1 \\ I_2 \end{bmatrix} = \begin{bmatrix} Y_L + Y_C & -Y_L \\ -Y_L & Y_L \end{bmatrix} \begin{bmatrix} v_t \\ v_b \end{bmatrix} \qquad (13)$$

where: $Y_L$ – transmission line admittance, $Y_c$ -local consumer's admittance.

By noting the admittance (Figure 3) $Y_1=Y_L+Y_C$ (own admittance of the network seen at generator terminals) and $Y_2=-Y_L$ (transfer admittance) and considering that

we are only interested by the generator's output current $I=I_1$, the equation which we will use is:

$$I = Y_1 v_t + Y_2 v_b \tag{14}$$

where: $Y_1=G_1+jB_1$ and $Y_2=G_2+jB_2$ ($G_1$, $G_2$ – conductance, $B_1$, $B_2$ – susceptance; adopted convention: $B_{1,2}<0$ - for inductance).

In this case, by projecting the current $I$ (equation (14)) on $d$ and $q$ axis, the relations that define the $i_d$ and $i_q$ currents will becomes:

$$i_d = G_1 v_d - B_1 v_q - G_2 v_b \sin(\delta) + B_2 v_b \cos(\delta) \tag{15}$$

$$i_q = G_1 v_q + B_1 v_d - G_2 v_b \cos(\delta) - B_2 v_b \sin(\delta) \tag{16}$$

Those relations ((15), (16)) represent the equations that define synchronous generator's connection to power system through a network that includes also a local consumer. Relations (15) și (16) can be brought to an equivalent form, resulting $v_d$ and $v_q$ voltages:

$$v_d = \frac{1}{B_1^2 + G_1^2}[i_d G_1 + i_q B_1 + (G_1 G_2 + B_1 B_2)v_b \sin(\delta) + (B_1 G_2 - B_2 G_1)v_b \cos(\delta)] \tag{17}$$

$$v_q = \frac{1}{B_1^2 + G_1^2}[i_q G_1 - i_d B_1 + (B_1 G_2 - B_2 G_1)v_b \sin(\delta) + (G_1 G_2 + B_1 B_2)v_b \cos(\delta)] \tag{18}$$

The reducing of non-linear model order (from six to four) can be accomplished by neglecting the transient effects in stator and the effects of rotor amortisation windings [6].

Considering these simplifying assumptions, the primary equation set of synchronous generator (1)…(10) is reduced to the following set of equations:

$$\frac{d}{dt}(\delta) = \omega_0 s \tag{19}$$

$$M \frac{d}{dt}(s) = -k_d s + T_m - T_e \tag{20}$$

$$T'_{d0} \frac{d}{dt}(e'_q) = v_f - (x_d - x'_d)i_d - e'_q \tag{21}$$

$$0 = v_d + \gamma_a i_d - x_q i_q \tag{22}$$

$$e'_q = v_q + \gamma_a i_q + x'_d i_d \tag{23}$$

$$T_e = e'_q i_q - (x'_d - x_q)i_d i_q \tag{24}$$

$$v_t^2 = v_d^2 + v_q^2 \qquad (25)$$

completed by equations (17),(18).

Linearising and converting the mathematical model from continuous to discrete time it is obtained a new linear model representing a 4$^{th}$ order transfer function (used only to design the self-tuning controller) [5].

$$H(z^{-1}) = z^{-1} \frac{B(z^{-1})}{A(z^{-1})} = z^{-1} \frac{b_3 z^{-3} + b_2 z^{-2} + b_1 z^{-1} + b_0}{a_4 z^{-4} + a_3 z^{-3} + a_2 z^{-2} + a_1 z^{-1} + 1} \qquad (26)$$

The performed simulations consider the case of the non-linear 6$^{th}$ order model of the synchronous generator (equations (1)…(10), completed with the connection network model – equations (17), (18)), as a process model integrated into the adaptive control system

## 3 Adaptive PI Fuzzy Control

There are many approach variants of the self-tuning problematic of fuzzy adaptive regulators. [2][3] A first method consists in the use of some scaling factors (on-line tunable) on the inputs and respectively, on the output of the fuzzy controller. Another method consists in the modification the membership function allure, through a suitable tuning of their parameter's values. The last mentioned method is the one used in the present paper.

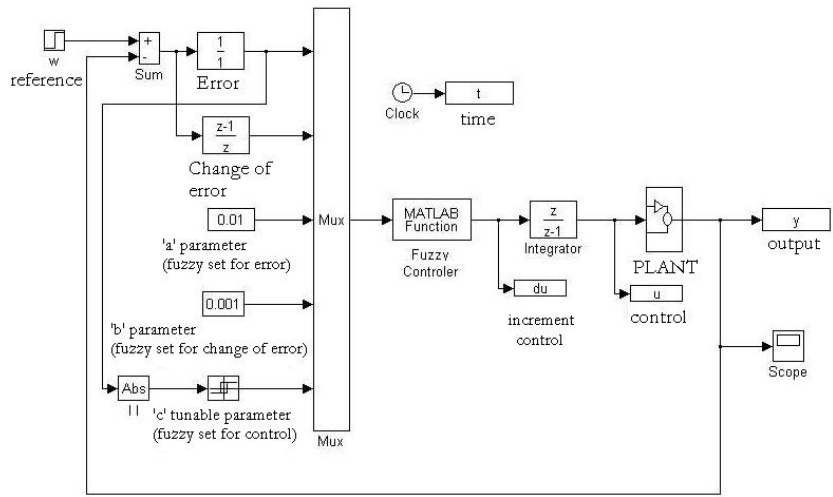

Figure 4
Simulink model of adaptive fuzzy PI control structure

In Figure 4 is presented the Simulink block diagram of the considered control structure. The output's error represents the main variable, on which relays the on-line adjusting of controller's parameter by the tuning algorithm. The implementation and simulation of control structure has been done with Matlab-Simulink, considering the case of a single tunable parameter (the strategy can be easily generalized also for all other parameters).

The on-line tuning algorithm of a parameter is implemented through a bipositional component, which briefly describes the following tasks (in this case, the 'c' parameter corresponding to the control's variable singletons):

- If the output's error remains in the specified values range, it is maintained a constant value c1 for 'c' parameter (experimental off-line determined value)

- If the output's error exceeds the values range, the 'c' parameter is tuned to a new value c2, and this value is maintained until the output's error decreases under a value (closed to zero).

## 4    Simulation Studies

The study cases showed the fact that the maintenance of 'c' parameter at the new *c2* tuned value (only until the output's error reaches again the specified values range) is not sufficient. In this case the tuning mechanism acts for a too short time period in order to be able to produce significant results in the controlled output. Based on this reason the β parameter has a very small chosen value. So, the tuning algorithm presents a similar behavior as a hysteresis relay element (Figure 5).

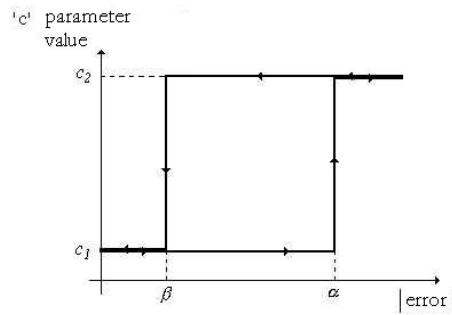

Figure 5
Evolution of tunable parameter

The following study considers as a plant the case of a synchronous generator connected to a power system. In Figure 6 is depicted the controlled output in both cases (step input response, perturbed at time moments 40 sec. and 60 sec.), with and without tuning (the time axis is scaled in seconds). All the figures have the abscissa axis scaled in time units (seconds). It is considered that any reference to an input or output variable is a deviation related to the steady state regime. The reference presents a variation of 0.05 p.u. at the moment t=20 s. At the moment t=40 s there can be noticed an increasing of mechanical torque (with 0.2 p.u.), and further, at t=60 s, a supplementary local consumer is connected to the system (G1 increase with 0.2 p.u.). Both events are regarded as perturbations for the considered system. The conducted studies proved that maintaining the value of 'c' parameter equal with $c_2$, only until the error reenter in the $[-\alpha,+\alpha]$ range is not sufficient, the readapted dynamic of output controller acts on a very short time interval, insufficient for produce benefic influences on controlled output. In this idea, the $\beta$ value is very small, close to zero, but not zero, because a zero error is a theoretical desiderate (especially if the system is stochastically perturbed).

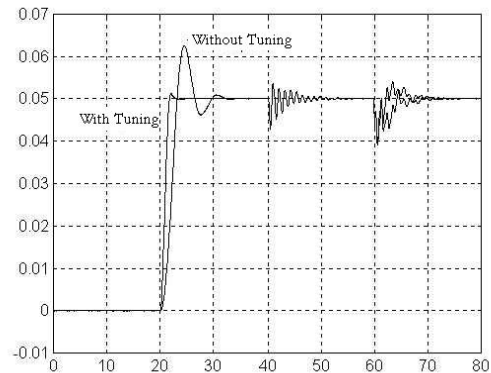

Figure 6
Controlled outputs

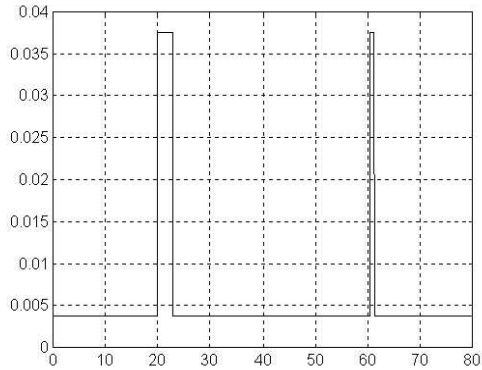

Figure 7
The tuning of 'c' parameter

There can be noticed the higher performances of the adaptive control structure (with on-line 'c' parameter tuning - Figure 7). Although the performances of the PID fuzzy control aren't presented, the mentioned PI fuzzy adaptive controller shows much better performances. For the considered plant, the self-tuning mechanism works especially in case of a reference modification. As the references variations grow, we concluded that the role of self-tuning mechanism is more necessary. Also, its role is welcomed when perturbations occur in the system.

**Conclusions**

In this paper an adaptive fuzzy control system based on a PI fuzzy controller is presented. There is proposed and validated an on-line self-tuning mechanism of controller's parameter. The performances of the proposed adaptive fuzzy control structure are compared with the one of a classical fuzzy control structure. The application has been developed related to the power system. All the study cases have been conducted through computer simulation by using Matlab-Simulink environment.